# 3D-printed terahertz subwavelength dual-core fibers with dense channel-integration

Haiyuan Ge, Haisu Li, Lu Jie, Jianshuai Wang, Yang Cao, Shaghik Atakaramians, *Senior Member, IEEE*, Yandong Gong, Guobin Ren, Li Pei

*Abstract*—Terahertz (THz) fiber that provides high-speed connections is one of the most essential components in THz communication systems. The emerging space-division-multiplexing technology is expected to increase the transmission capacity of THz communications. A promising candidate to achieve that is integrating multiple channels in a compact THz multi-core fiber system. Here, we propose and experimentally demonstrate a THz subwavelength rectangular dielectric dual-core fiber structure, where two identical cores can be densely integrated, thanks to the polarization-maintaining feature of the rectangular fiber. Different configurations of the fiber structure, including the placements, core-spacings, and polarization states of two fiber cores, are comprehensively investigated to improve the channel isolation. Numerical simulations show that the fractional power in core of fiber mode has a dominant effect on inter-core coupling performance. Moreover, we design the core size (1 mm x 0.5 mm) slightly less than the WR5.1 waveguide (1.295 mm x 0.6475 mm) so that the fiber can be conveniently connected with the WR5.1 flange port with mode excitation efficiencies up to 62.8%. A cost-efficient dielectric 3D printing technique is employed for rapid fabrications of dual-core fibers as well as corresponding polymer flange structures that offer solid integration between the fiber samples and the WR5.1 port. Experimental measurements of dual-core fibers demonstrate that a 4-mm core-spacing (less than three times of the operation wavelengths over a frequency range of 0.17-0.21 THz) is sufficient to support robust dual-channel propagation with channel isolation values more than 15 dB, which are consistent with the theoretical and numerical results. This work provides a densely integrated dual-core fiber system with low fabrication cost and practical connection to WR5.1 flange, holding exciting potentials for high-capacity THz space-division-multiplexing communication systems.

This work was supported in part by the National Natural Science Foundation of China under Grants 62075007, 62275011, and 62205100; and in part by Fundamental Research Funds for the Central Universities under Grant 2022JBZY004. *(Corresponding author: Haisu Li).*

Haiyuan Ge, Haisu li, Lu Jie, Jianshuai Wang, Guobin Ren and Li Pei are with the Key Laboratory of All Optical Network and Advanced Telecommunication Network of EMC, Institute of Lightwave Technology, Beijing Jiaotong University, Beijing 100044 China (e-mail: 22120053@bjtu.edu.cn; lihaisu@bjtu.edu.cn; 21120067@bjtu.edu.cn; wangjsh@bitu.edu.cn; gbren@bjtu.edu.cn; lipei@bjtu.edu.cn).

Yang Cao is with the Center for Advanced Laser Technology, Hebei University of Technology (e-mail: yang.cao@hebut.edu.cn).

Shaghik Atakaramians is with the Terahertz Innovation Group, School of Electrical Engineering & Telecommunications, UNSW, Sydney, NSW 2052 Australia (email: s.atakaramians@unsw.edu.au).

Yandong Gong is with the School of Instrument Science and Optoelectronics Engineering, Beijing Information Science and Technology University (email: eydgong@bistu.edu.cn).

## I. INTRODUCTION

THZ radiation (0.1-10 THz), sandwiched between microwave and infrared light waves, has attracted tremendous attention in various applications from scientific research to our daily-life. In the field of communications, the shift of carrier frequencies to the THz band is inevitable so as to meet the bandwidth requirements of next-generation wireless systems [1], [2]. THz waves have been demonstrated to be suitable for both line-of-sight and non-line-of-sight communications [3]. Nevertheless, THz wireless communications face several challenges, such as atmospheric absorption in weather conditions such as rain, fog, and snow [4], and the high directionality of the THz waves resulting in the requirement for precise alignment of the transmitter and receiver [5], which would degrade the performance and reliability of THz transmission links. THz wired communication, on the contrary, can effectively compensate for those issues of wireless counterparts because THz waveguides are able to propagate modes spanning intricate geometrical paths, presenting a secure and stable propagation environment. Moreover, THz waveguides could provide reliable coupling and connection to the transmitter and receiver for both static and dynamic applications. Many demonstrations of wired communication links based on THz waveguides have been reported in recent years, which can be classified into THz planar waveguides and fiber waveguides according to the waveguide structures [6]. THz planar silicon waveguides, such as photonic crystal waveguides [7], [8], topological planar waveguides [9], [10], and effective dielectric clad dielectric waveguides [11], are usually suitable for on-chip THz communications with data-rate up to 36 Gbps. For long-distance transmission (up to 10 m), THz flexible fiber waveguides including solid-core waveguides [12], [13], hollow-core waveguides [14], [15], and suspended-core waveguides [16] are preferable. Assisted by the high-order modulation technique, the data rate of the THz fiber-based communication system has reached 352 Gbps [17], [18].

However, current THz wired communications usually exploit single-channel fibers and the improvement of the transmission capacity largely depends on high-order modulation formats. Inspired by the space-division-multiplexing technique that has been intensively applied for optic-communication regimes, a THz fiber integrating multiple cores could further enhance the transmission capacity from the physical layer. A fundamental structure is a dual-core fiber. Recently, Li et al. have investigated the coupling between two parallel-placed polytetrafluoroethylene circular dielectric fibers over 0.09-0.11 THz [19]. Measured results show that for coupling lengths of 50 mm and 100 mm, the coupling power can be kept under -30 dB for both horizontal and vertical polarization states at a core-spacing of 9 mm. Moreover, several polarization-sensitive THz devices, such as polarization beam splitters [20], [21], [22], [23], [24], [25], directional couplers [26], [27], [28], [29], [30], [31], [32] and sensors [33], [34], [35], have also been realized based on the dual-core fiber structures. Taking a variable coupler consisting of two 60-mm-long parallel fibers as an example, experiments show that by adjusting the distance (0-3 mm) between the main and coupled fibers, the power between both fibers could be exchanged [26]. Generally, there exists a tradeoff between the channel crosstalk and the integration density of channels. For square- [26], [27] and circular-core [19] dielectric fibers that are polarization-independent, the polarization states could be changed during propagation, which would result in unexpected channel coupling and require large core-spacing. Another practical challenge is the coupling between the fiber and the THz source/detector, where additional components are usually required, such as lenses and horn couplers [15], [18], [27]. This will increase the cost and complexity of the THz system.

In order to improve the integration density of the dual-core fiber system and the excitation efficiency with commonly used flanges at THz bands, this work proposes subwavelength rectangular dielectric dual-core fibers with high birefringence, which are fabricated using fused-deposition-modeling based 3D printers. In the following content, we first investigate the coupling effects of the polarization-maintaining dual-core fiber systems, including the placement, core-spacing, and polarization states of two fiber cores. Next, 3D-printed dual-core fibers with various coupling region lengths and core-spacings are measured and analyzed. Both theoretical and experimental results confirm that the channel isolation of dual-core fiber with a 4-mm core-spacing is over 15 dB, which could be applied for THz space-division-multiplexing communication systems. Furthermore, the subwavelength rectangular-core fibers can be easily connected with THz WR5.1-port devices assisted with the 3D-printed dielectric flanges.

## II. FIBER DESIGN FROM SINGLE-CORE TO DUAL-CORE

### A. Subwavelength rectangular single-core fiber

Before moving to the dual-core fiber, we first design a dielectric fiber that has a single subwavelength rectangular core with a cross-sectional size of 1 mm x 0.5 mm (target frequency ranging from 0.17 THz to 0.21 THz, or wavelengths of 1.43-1.76 mm), as shown in Fig. 1(a), where the THz wave is transmitted along the z-direction. On one hand, the rectangular core has been demonstrated to support high birefringent transmission [36], [37], which is expected to improve the integration density of dual-core structures. On the other hand, the core dimension is slightly smaller than the standard WR5.1-waveguide port (1.295 mm x 0.6475 mm), so that the fiber would be directly connected with the WR5.1 flange port. We carry out finite-element-method-based simulations of the single-core fiber using the commercial software COMSOL [38]. In the numerical models, the dielectric is acrylonitrile butadiene styrene (ABS) – a commonly used 3D-printable polymer, the refractive index over 0.17-0.21 THz of which is in accordance with the experimental characterization from [39]. Fig. 1(b) shows the fractional power in core of the two orthogonally polarized modes (denoted as *a-pol.* and *b-pol.*, the electric fields of which are polarized along the x-axis and y-axis, respectively). As the frequency increases, the corresponding wavelength decreases, so more power is confined in core. This is because the subwavelength dielectric fiber has a leaky mode character, where part of the power is distributed outside the subwavelength fiber [36]. Due to the rectangular core shape, the *a-pol.* mode has a larger fractional power in core than that of the *b-pol.* mode. Therefore, the effective refractive index of *a-pol.* mode is higher than that of *b-pol.* mode, as shown in Fig. 1(c). Over 0.17-0.21 THz, the birefringence values between *a-* and *b-pol.* modes are higher than 6 x 10$^{-2}$.

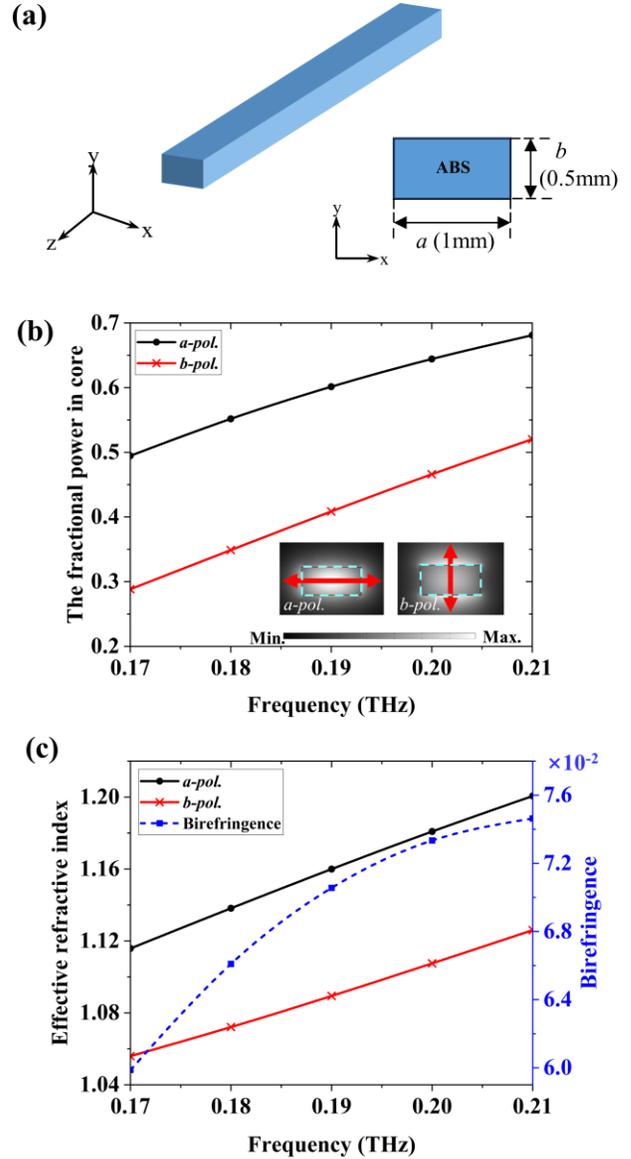

Fig. 1. (a) Schematic of the rectangular fiber. (b) The fractional power in core as a function of frequency; the inset shows the normalized electric fields of the *a-pol.* and *b-pol.* modes at 0.2 THz. (c) Effective refractive index and modal birefringence of the *a-pol.* and *b-pol.* modes as a function of frequency.

TABLE I
COUPLING COEFFICIENTS AND COUPLING LENGTHS FOR 10 CONFIGURATIONS AT 0.2 THZ

| | Configurations of dual-core fiber | | | Coupling coefficient $C_{AB}$ | Coupling coefficient $C_{BA}$ | Coupling length ($L_{AB}$ / cm) | Coupling length ($L_{BA}$ / cm) |
|---|---|---|---|---|---|---|---|
| | core A | core B | | | | | |
| 1 | HP, *a-pol.* | HP, *a-pol.* | 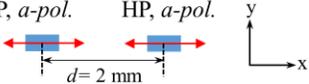 | 15.06 | 15.06 | 10.43 | 10.43 |
| 2 | HP, *b-pol.* | HP, *b-pol.* | 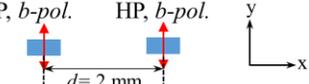 | 20.04 | 20.04 | 7.84 | 7.84 |
| 3 | HP, *a-pol.* | HP, *b-pol.* | 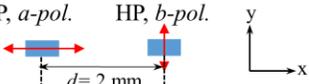 | 1.764 | 3.284 | 89.01 | 47.83 |
| 4 | VP, *a-pol.* | VP, *a-pol.* | 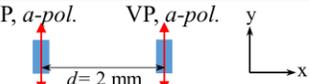 | 6.523 | 6.523 | 24.08 | 24.08 |
| 5 | VP, *b-pol.* | VP, *b-pol.* | 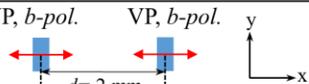 | 21.38 | 21.38 | 7.34 | 7.34 |
| 6 | VP, *a-pol.* | VP, *b-pol.* | 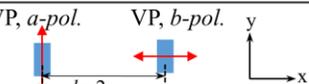 | 4.051 | 0.7308 | 38.78 | 214.92 |
| 7 | HP, *a-pol.* | VP, *a-pol.* | 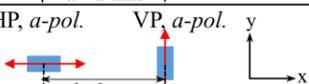 | 0.5876 | 2.139 | 267.32 | 73.43 |
| 8 | HP, *b-pol.* | VP, *b-pol.* | 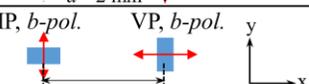 | 5.706 | 2.408 | 27.53 | 65.21 |
| 9 | HP, *a-pol.* | VP, *b-pol.* | 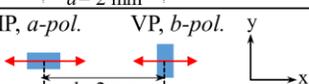 | 29.72 | 9.828 | 5.28 | 15.98 |
| 10 | HP, *b-pol.* | VP, *a-pol.* | 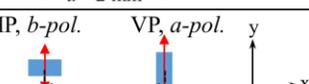 | 6.811 | 20.77 | 23.06 | 7.56 |

*B. Coupling of dual-core fiber structures*

Here we employ two identical subwavelength rectangular cores (denoted as core A and core B) to design the dual-core fiber structure, the inter-core coupling can be estimated based on coupled mode theory, which is expressed as follows [40]:

$$C_{AB} = \frac{2\pi f \varepsilon_0 \times \int (n^2 - 1^2) \mathbf{E}_A^* \cdot \mathbf{E}_B \cdot d\mathbf{S}}{\int \mathbf{u}_z \cdot (\mathbf{E}_A^* \times \mathbf{H}_A + \mathbf{E}_A \times \mathbf{H}_A^*) \cdot d\mathbf{S}} \quad (1)$$

where $\mathbf{E}_A$ and $\mathbf{H}_A$ are the electric and magnetic fields of core A, $\mathbf{E}_B$ is the electric field of core B, $n$ is the refractive index of the fiber core, $\varepsilon_0$ is the vacuum permittivity, and $f$ is the operation frequency. $C_{AB}$ is the overlap of the mode fields of core A and core B in the cross-section of core A, indicating the power transferred from core A to core B. The coupling length $L_{AB}$ can thus be calculated as:

$$L_{AB} = \frac{\pi}{2C_{AB}} \quad (2)$$

To achieve a dense integration of two fiber cores, the isolation between two channels should be improved, i.e., lower $C_{AB}$ or higher $L_{AB}$ values. In the following contents, we investigate the inter-core coupling of 10 sets of dual-core fiber configurations, according to placements and the polarization states of each core, as shown in Table I. For either the long- or short-side of the rectangular core along the x-axis, the corresponding configuration is denoted as horizontal placement (HP) or vertical placement (VP), respectively. Moreover, the red arrows indicate the electric vector of the *a-pol.* and *b-pol.* modes. Simulation results of inter-core coupling performance for the 10 configurations at 0.2 THz are summarized in Table I, where the core-spacing (*d*) keeps invariant (*d* = 2 mm).

First, for dual-core fiber configurations 1, 2, and 3 with

HP, cores guiding the *a-pol.* mode offers higher isolation values than the *b-pol.* mode because more power is confined in the core of the *a-pol.* mode. Moreover, we find that the dual-core structure with two orthogonally polarized modes (i.e., configuration 3) could significantly reduce the inter-core coupling, thanks to the polarization-maintaining nature of the subwavelength rectangular core. Second, similar phenomena can be seen for dual-core configurations 4, 5, and 6 with VP. We note that, although both configurations 1 and 4 guide *a-pol.* modes, VP has lower inter-core coupling than that of HP. The reason is that the two cores are placed along the x-axis, so the two modes with polarization state along the y-axis (whatever *a-pol.* or *b-pol.* mode) provide high channel isolations. Third, we study the dual-core fiber structure consisting of both HP and VP cores, i.e., configurations 7, 8, 9, and 10. We highlight that configuration 7 could support the highest channel isolation performance in Table I, since two strongly confined *a-pol.* modes are orthogonal to each other. From the above discussions, we conclude that the confinement of mode plays a critical role in power exchange between two cores. Nonetheless, it is worthwhile noting that, a large dielectric core size would result in an enhancement of both channel isolation and transmission loss [36]. Therefore, the dimension of the subwavelength core should be carefully designed to balance the trade-off between the channel isolation and fiber loss.

### III. MODE EXCITATION, FABRICATION, AND EXPERIMENT OF DUAL-CORE FIBER

*A. Mode excitation*

To improve the integration of THz systems, THz sources and detectors with flange ports are preferable in contrast to the photoconductive antenna commonly used in THz time-domain spectroscopy systems. For the target frequency range of 0.17-0.21 THz, a THz multiplier with a WR5.1 waveguide port is applied in this work to excite the dual-core fiber, where the subwavelength core can be readily inserted inside the WR5.1 port. We evaluate the overlap integral (i.e., the excitation efficiency) between the *a(b)-pol.* modes of the proposed subwavelength rectangular dual-core fiber and the $TE_{10}$ mode of metallic WR5.1 waveguide, using the following equation [41]:

$$overlap = \left| Re\left[ \frac{\left(\int_{S_1} \boldsymbol{E}_1 \times \boldsymbol{H}_2^* \cdot d\boldsymbol{S}\right)\left(\int_{S_1} \boldsymbol{E}_2 \times \boldsymbol{H}_1^* \cdot d\boldsymbol{S}\right)}{\int_{S_1} \boldsymbol{E}_1 \times \boldsymbol{H}_1^* \cdot d\boldsymbol{S}} \right] \frac{1}{Re\left(\int_{S_1} \boldsymbol{E}_2 \times \boldsymbol{H}_2^* \cdot d\boldsymbol{S}\right)} \right| \quad (3)$$

where $\boldsymbol{E}_1$ and $\boldsymbol{H}_1$ represent the electric field and magnetic field of the rectangular fiber, respectively, $\boldsymbol{E}_2$ and $\boldsymbol{H}_2$ represent the electric field and magnetic field of the WR5.1 waveguide, respectively, and the integration region $S_1$ is a square with a side length of 15 mm where 99.99% power of the rectangular subwavelength fiber resides. Fig. 2 shows the mode excitation efficiencies of the *a-pol.* and *b-pol.* modes of dual-core fiber over 0.17-0.21 THz. Since the electric field of *b-pol.* mode is polarized along the same direction of the $TE_{10}$
mode in metallic WR5.1 waveguide, the excitation efficiency of *b-pol.* mode could be up to 62.8%, as the red curve shown in Fig. 2. On the contrary, the excitation efficiency of the *a-pol.* mode is below $2.2 \times 10^{-5}$. We should note that a mode converter could be employed to convert the $TE_{10}$ mode to the $TE_{01}$ mode of the WR5.1 waveguide to efficiently excite the *a-pol.* mode.

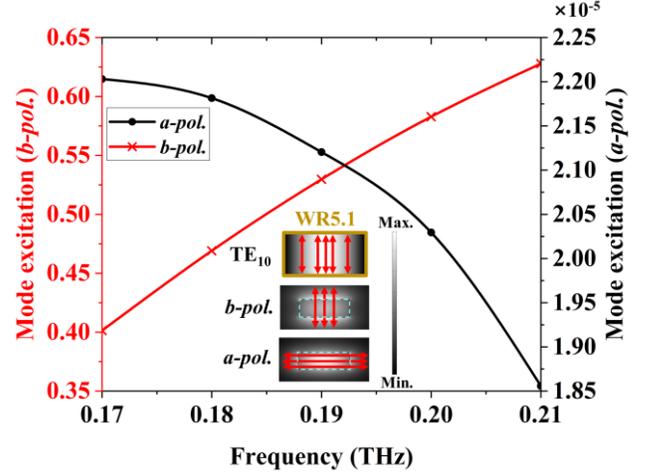

Fig. 2. Mode excitation for *a-pol.* and *b-pol.* modes in rectangular fiber with $TE_{10}$ mode in metallic WR5.1 waveguide as a function of frequency. The inset shows the mode normalized electric fields of the $TE_{10}$ mode, *a-pol.* mode and *b-pol.* mode, and the red arrows indicate the electric vectors.

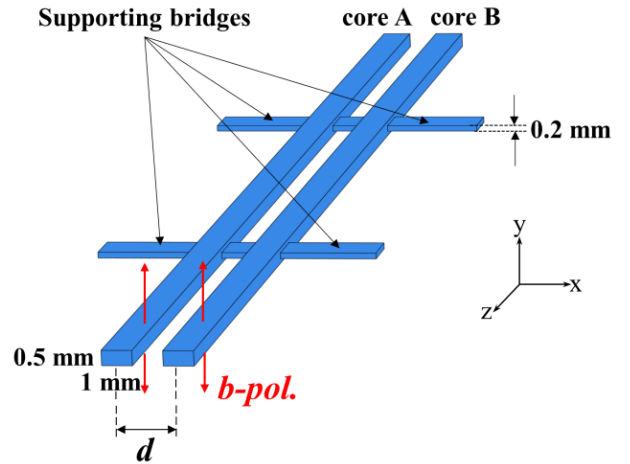

Fig. 3. Schematic of the dual-core fiber structure where the red arrows indicate the electric vectors.

*B. Fabrication and experiment setup*

In practice, a supporting bridge is required to implement the proposed fiber structure consisting of two subwavelength cores. Additionally, because of the efficient excitation of the *b-pol.* mode of the fiber using the WR5.1 port, configurations of 2, 5, and 8 shown in Table I are promising candidates in the context of fabrications. However, for configurations 5 and 8, the additional supporting bridge would increase the losses of *b-pol.* mode, since the THz power would be propagated along the lossy dielectric bridges. Therefore, we choose configuration 2 as the dual-core fiber structure for later fabrications. Fig. 3 shows the design structure of configuration 2, which includes two supporting bridges with a thickness of 0.2 mm. The overall length of two supporting bridges is designed to match the inner diameter of the WR5.1 flange (9.53 mm), which would facilitate a solid connection of the dual-core fiber to the flange (the details will be shown in Fig. 5).

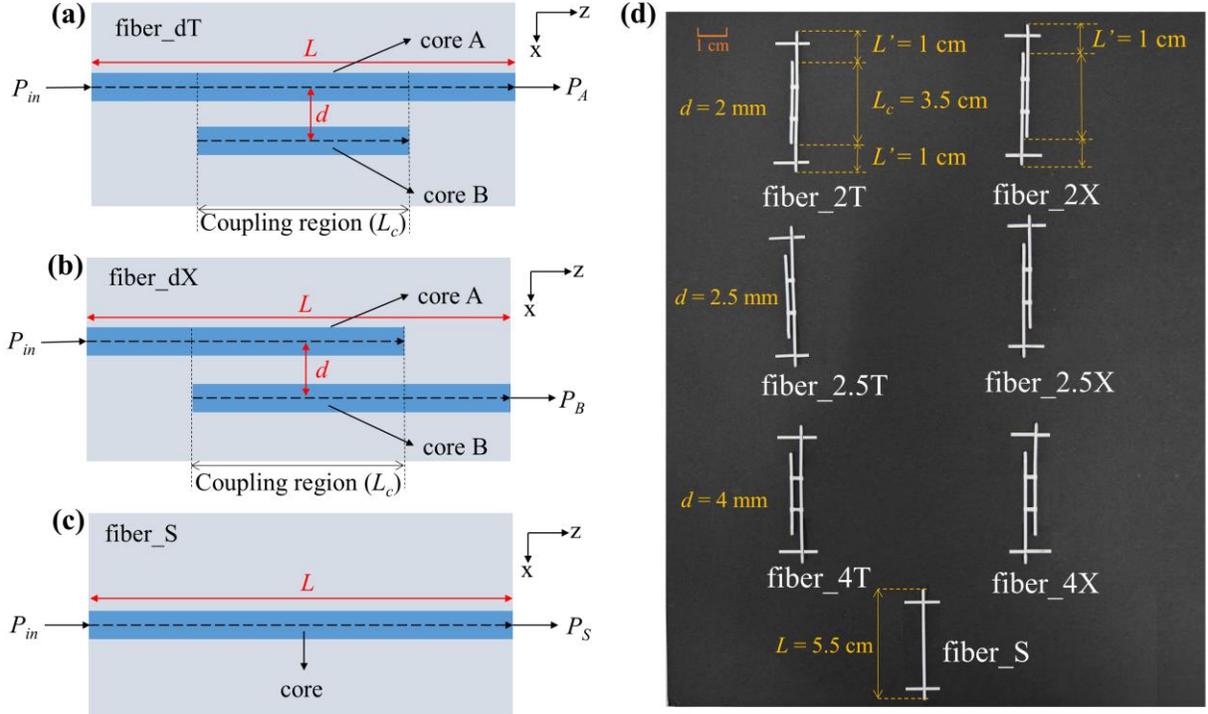

Fig. 4. (a) The schematic structure of fiber$_{dT}$, where $P_A$ is the output power of core A. (b) The schematic structure of fiber$_{dX}$, where $P_B$ is the output power of core B. (c) The schematic structure of fiber$_S$, where $P_S$ is the output power of a single-core. (d) 3D-printed samples of both dual-core ($L_c$ = 3.5 cm, $d$ = 2 mm, 2.5 mm and 4 mm) and single-core ($L$ = 5.5 cm) fibers.

TABLE II
3D-PRINTED SAMPLES OF DUAL-CORE AND SINGLE-CORE FIBERS

|  | fiber$_{dT}$ | fiber$_{dX}$ | fiber$_S$ |
|---|---|---|---|
| dual-core ($d$ = 2 mm) | $L$ = 5.5 cm, 7 cm, 8.5 cm | $L$ = 5.5 cm, 7 cm, 8.5 cm |  |
| dual-core ($d$ = 2.5 mm) | $L$ = 5.5 cm, 7 cm, 8.5 cm | $L$ = 5.5 cm, 7 cm, 8.5 cm |  |
| dual-core ($d$ = 4 mm) | $L$ = 5.5 cm, 7 cm, 8.5 cm | $L$ = 5.5 cm, 7 cm, 8.5 cm |  |
| single-core |  |  | $L$ = 5.5 cm, 7 cm, 8.5 cm |

A Pro 3 Plus printer from Raise 3D is employed in our work to fabricate the dual-core fibers with total lengths $L$ = 5.5 cm, 7 cm, and 8.5 cm, and the corresponding lengths of coupling regions are $L_c$ = 3.5 cm, 5 cm, and 6.5 cm, respectively. For each length, the dual-core fibers have three core-spacings, including $d$ = 2 mm, 2.5 mm, and 4 mm. In order to evaluate the coupling between core A and core B (note that only core A is excited with an input power of $P_{in}$ in the following contents), two structures of the dual-core fiber (denoted as fiber$_{dT}$ and fiber$_{dX}$, where $d$ indicates the core-spacing with a unit of mm) are designed to separately measure the power through each core ($P_A$ and $P_B$), where the $L_c$ lengths of two structures keep the same, as shown in Fig. 4(a) and (b). Moreover, three single-core fibers with identical total lengths $L$ [denoted as fiber$_S$, as shown in Fig. 4(c)] are also 3D-printed for comparison purposes. In a nutshell, 21 fiber samples are 3D-printed, as summarized in Table II. Fig.

4(d) shows the photos of 3D-printed fiber samples with $L_c$ = 3.5 cm and $d$ = 2 mm, 2.5 mm, and 4 mm.

Fig. 5 shows the schematic and the actual setup of the experiment. We first generate signals ranging from 14.2 GHz to 17.5 GHz using a signal generator (Ceyear, 1435D), which are then amplified 12 times using a multiplier (Ceyear, 82406C), producing THz signals over 0.17-0.21 THz. The fiber sample under test is connected to a WR5.1 waveguide, which is linked to the multiplier, as shown in Fig. 5. A power meter (Ceyear, 2438CA) equipped with a WR5.1-port probe receives the THz power after propagating through the fiber sample. Two 3D-printed dielectric flanges (the sample photo is presented in the inset of Fig. 5) are employed at both input and output ends of the fiber, which are respectively integrated with metallic flanges of the WR5.1 waveguide and the THz probe, as shown in Fig. 5.

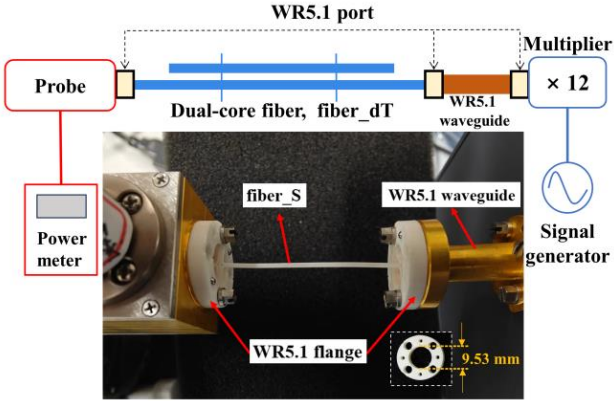

Fig. 5. The schematic and the actual setup of the experiment. The inset presents a 3D-printed flange.

IV. EXPERIMENTAL RESULTS AND DISCUSSION

In this section, we first use the experimental setup of Fig. 5 to measure single-core fibers and characterize the propagation loss. Next, we measure the dual-core fibers to investigate the effects of core-spacing and coupling region length on the transmission performance and integration density.

*A. Single-core fiber transmission measurement*

First, we perform the transmission measurement of single-core fibers [i.e., the fiber$_S$ shown in Fig. 4(d)] with $L$ = 5.5 cm, 7 cm, and 8.5 cm. The transmission of three single-core fibers (normalized to the power of only a WR5.1 waveguide under test) is presented in Fig. 6(a). The transmission values of *b-pol.* mode decrease with the fiber length and the frequency due to an improvement of the factional power in core [see Fig. 1(b)]. Next, we adopt the cut-back approach to characterize the fiber loss. As shown in Fig. 6(b), the purple solid curve shows the averaged value of the experimental losses, and the error bars indicate the standard deviation of the measurements. Moreover, the averaged value of the losses is linearly fitted [the fitting function is presented in Fig. 6(b)], which varies from 0.778 dB/cm to 2.42 dB/cm within 0.17-0.21 THz, as the red solid line shown in Fig. 6(b). We suggest that using other low-loss materials (e.g., Zeonex [42]) would further reduce the transmission loss. Moreover, we observe the transmission curves exhibit resonances spaced at approximately 4 GHz in Fig. 6(a). This is because of the leaky mode of the subwavelength-core fiber, where a portion of the power is distributed outside the subwavelength core. While the port walls of the WR5.1 waveguide and THz probe are metallic, leading to a linear resonant cavity (length of 5.5 cm) for the THz waves guided along the surface. The resonance spacing is in accordance with the free-spectral-ranges (values are 4.15 GHz, 3.97 GHz, and 3.82 GHz at 0.17 THz, 0.18 THz, and 0.19 THz, respectively) [43].

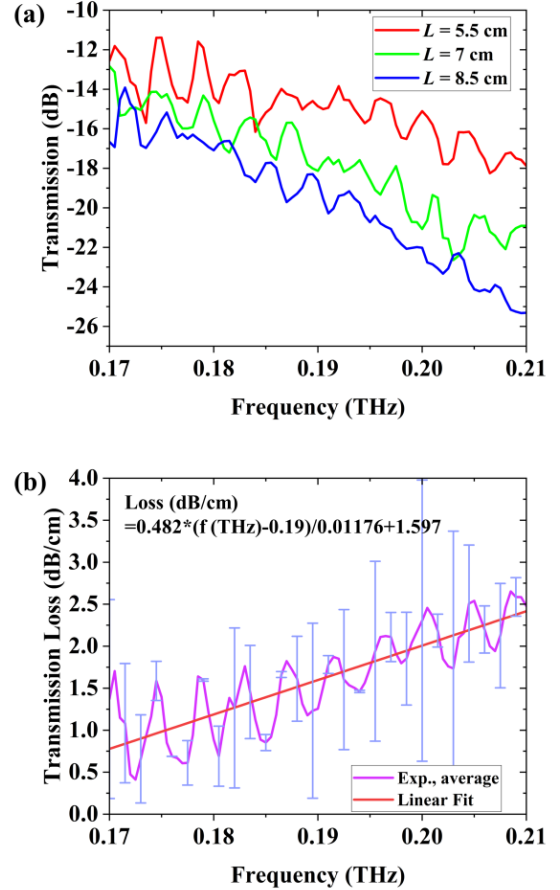

Fig. 6. (a) The transmission of *b-pol.* mode for single-core fibers as a function of frequency. (b) Single-core fiber transmission losses as a function of frequency. The purple solid curve shows the experimental results, the red solid curve shows the fitting results and the error bars indicate the standard deviation of the measurements.

*B. Dual-core fiber transmission measurement*

Here, we experimentally investigate the coupling characteristics of dual-core fibers, the measured power values of two cores are normalized to the power of the WR5.1 waveguide over 0.17-0.21 THz. First, we study dual-core fibers with $d$ = 2 mm. Fig. 7(a)-(c) show the experimentally measured normalized transmission for three coupling regions of $L_c$ = 3.5 cm, 5 cm, and 6.5 cm, where $T_S$ (the red solid curve) is the transmission of the single-core fiber, $T_A$ (the green solid curve) is the transmission of core A, and $T_B$ (the blue solid curve) is the transmission of core B. Fig. 7(d) shows the theoretical values of the coupling coefficient and coupling length. Moreover, we plot the normalized analytical power curves of the dual-core fiber based on coupled mode theory as shown in Fig. 7(e), where core A is excited, $P_A$ and $P_B$ are the output power of core A and core B, respectively, and Z denotes the transmission distance. Fig. 7(f) shows the numerical results based on the finite-difference-time-domain method to depict the normalized power distributions of dual-core fibers at 0.19 THz. From Fig. 7(a)-(c) ($d = 2$ mm), we observe power exchange between core A and core B, which is influenced by both frequency and $L_c$. In Fig. 7(a), experimental results indicate that the output powers of core A and core B are approximately equal at around 0.198 THz. At frequencies below 0.198 THz, the majority of power is transferred from core A to core B ($T_B > T_A$), whereas at frequencies above 0.198 THz, relatively more power remains in core A ($T_A > T_B$). Additionally, when $L_c = 6.5$ cm [Fig. 7(c)], a notable difference between $T_A$ and $T_B$ at 0.19 THz is observed, which can be attributed to the fact that the coupling region of 6.5 cm is relatively close to the theoretical coupling length of 6 cm, as labeled in Fig. 7(d). In other words, at $L_c = 6.5$ cm, the power from core A is almost transferred to core B, which is further validated by the theoretical and numerical results presented in Fig. 7(e) (blue curves) and Fig. 7(f) for $Z$ = 6.5 cm.

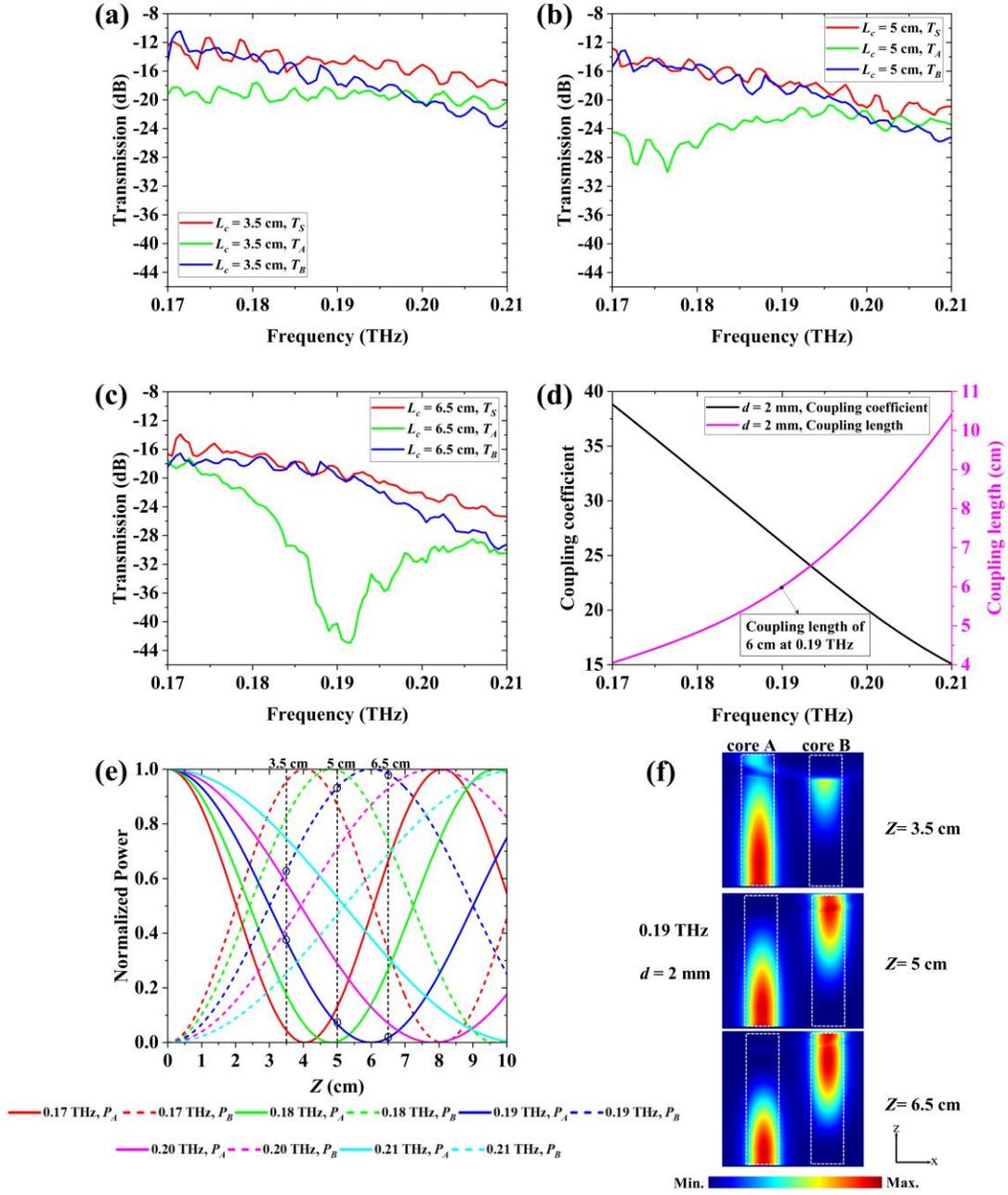

Fig. 7. Experimental results: the transmission of fibers with $d = 2$ mm, where $T_S$ (red solid curves) is the transmission of the single-core fiber, $T_A$ (green solid curves) is the transmission of core A, $T_B$ (blue solid curves) is the transmission of core B for (a) $L_c = 3.5$ cm, (b) $L_c = 5$ cm, and (c) $L_c = 6.5$ cm. (d) Simulation results: coupling coefficients and coupling lengths as a function of frequency for $d = 2$ mm. (e) Analytical results: normalized power in core A (solid curve, $P_A$) and core B (dotted curve, $P_B$) as a function of transmission distance (Z) for $d = 2$ mm. (f) Numerical results: normalized power distributions of core A and core B for $Z = 3.5$ cm, 5 cm, and 6.5 cm ($d = 2$ mm, 0.19 THz).

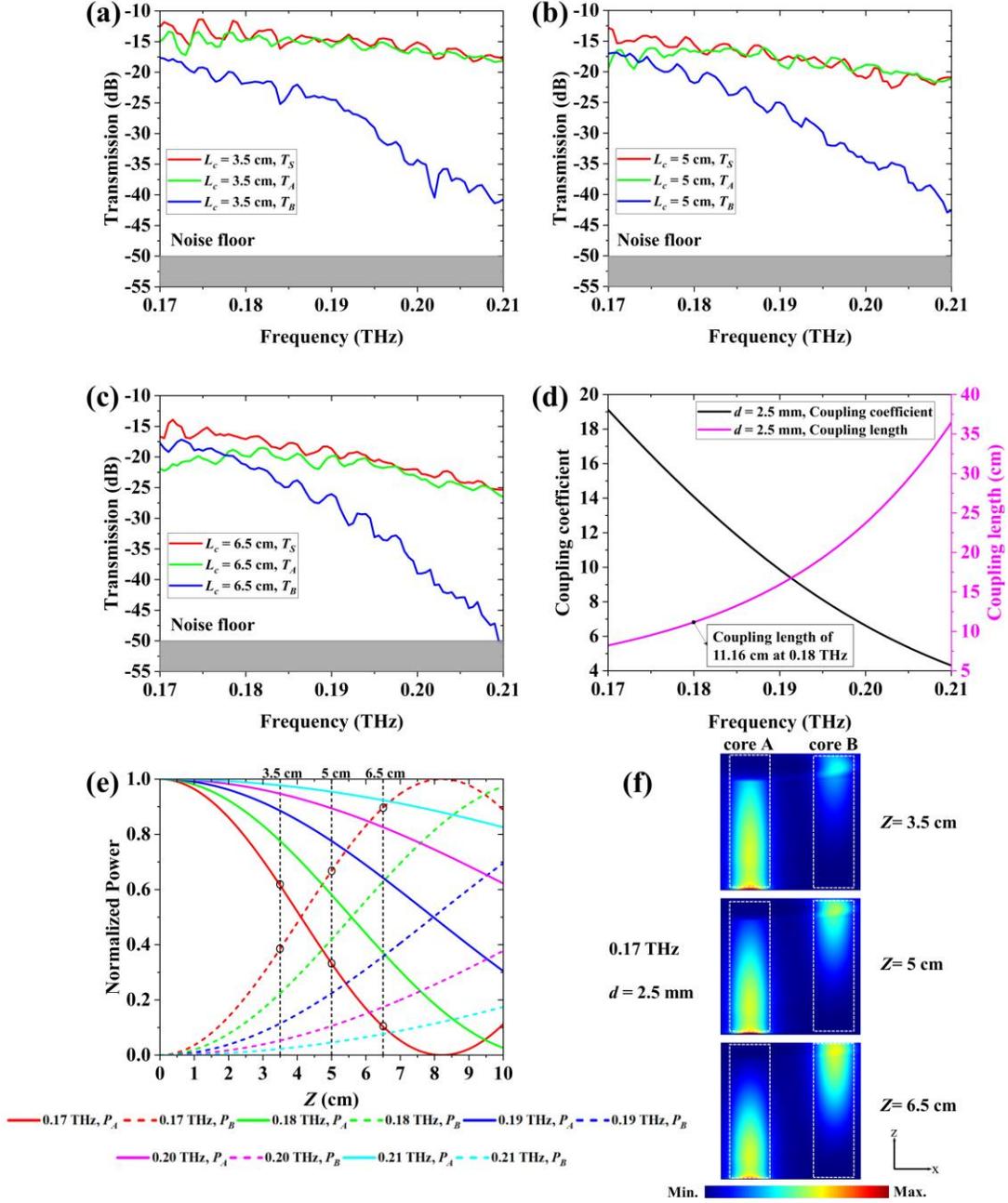

Fig. 8. Experimental results: the transmission of fibers with $d$ = 2.5 mm, where $T_S$ (red solid curves) is the transmission of the single-core fiber, $T_A$ (green solid curves) is the transmission of core A, $T_B$ (blue solid curves) is the transmission of core B for (a) $L_c$ = 3.5 cm, (b) $L_c$ = 5 cm, and (c) $L_c$ = 6.5 cm. (d) Simulation results: coupling coefficients and coupling lengths as a function of frequency for $d$ = 2.5 mm. (e) Analytical results: normalized analytical power in core A (solid curve, $P_A$) and core B (dotted curve, $P_B$) as a function of transmission distance ($Z$) for $d$ = 2.5 mm. (f) Numerical results: numerical results of normalized power distributions of core A and core B for $Z$ = 3.5 cm, 5 cm, and 6.5 cm ($d$ = 2.5 mm, 0.17 THz).

Next, we extend the core-spacing to 2.5 mm, the results of which are presented in Fig. 8. In Fig. 8(a)-(c), the power exchange between core A and core B only occurs at low frequencies (0.17-0.18 THz). As the coupling region increases, more power is transferred from core A to core B. Theoretical and numerical results respectively shown in Fig. 8(e) (red and green curves) and Fig. 8(f) (normalized power distributions) verify the power exchange between core A and core B at low frequencies. For frequencies beyond 0.18 THz, $T_S$ and $T_A$ are roughly the same since the theoretical coupling length (11.16 cm at 0.18 THz [labeled in Fig. 8(d)]) has exceeded the coupling regions designed for the dual-core samples with $d$ = 2.5 mm. Furthermore, due to the limited dynamic range of the experimental system, any transmissions less than -50 dB are considered to be noises.

Finally, we investigate the characteristics of the dual-core fiber with $d$ = 4 mm. From Fig. 9(a)-(c), we observe extremely weak couplings between core A and core B for all three fibers with different coupling regions, since the theoretical coupling length is 63.41 cm at 0.17 THz, as labeled in Fig. 9(d). The channel isolations (the difference between $T_A$ and $T_B$) of dual-core fibers are higher than 15 dB over 0.17-0.21 THz. Moreover, the isolation values increase with higher frequencies, because of the stronger mode confinement [see Fig. 1(b)].

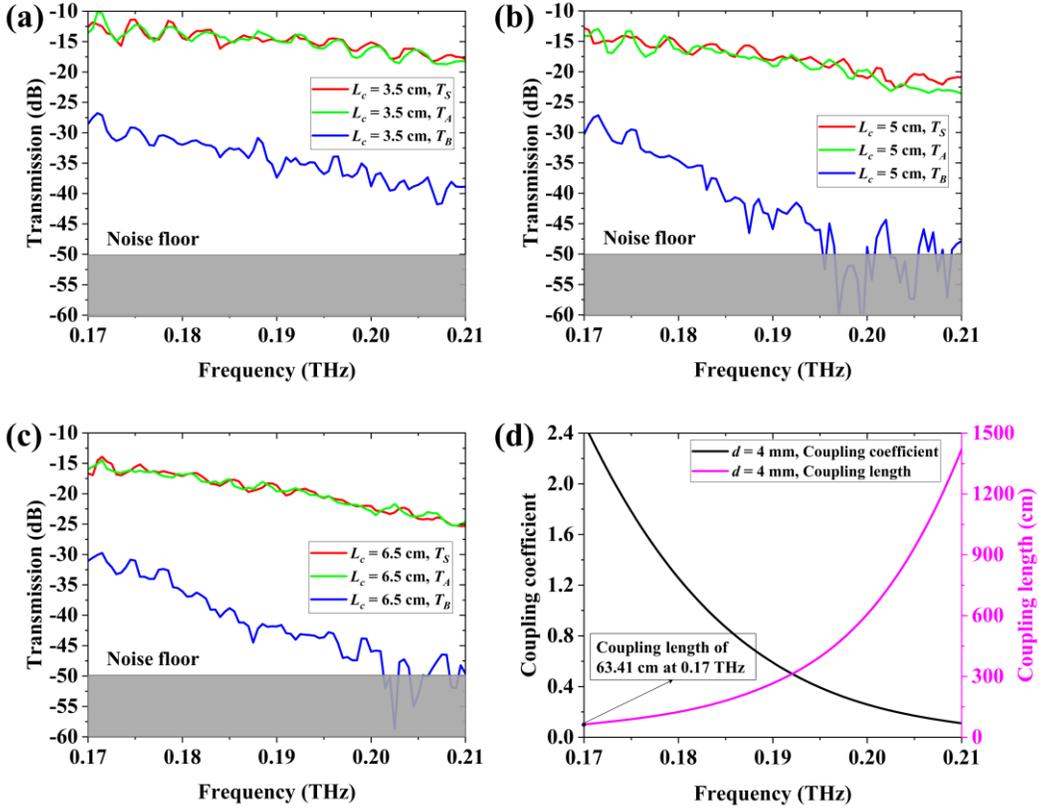

Fig. 9. Experimental results: the transmission of fibers with $d = 4$ mm, where $T_S$ (red solid curves) is the transmission of the single-core fiber, $T_A$ (green solid curves) is the transmission of core A, $T_B$ (blue solid curves) is the transmission of core B for (a) $L_c = 3.5$ cm, (b) $L_c = 5$ cm, and (c) $L_c = 6.5$ cm. (d) Simulation results: coupling coefficients and coupling lengths as a function of frequency for $d = 4$ mm.

TABLE III
PERFORMANCE COMPARISON OF DUAL-CORE FIBERS

| Core shape of dual-core fibers | Polarization sensitivity | Normalized core-spacing ($d/\lambda$) | Channel isolation | Fiber excitation and connection |
|---|---|---|---|---|
| Rectangular-shape (this work) | Yes | 2.53 | >15 dB | WR5.1-port flanges |
| Circular-shape ([19]) | No | 3 | Unmeasured | Additional adaptors and loads |

According to the measured results, the slight variations in core-spacing (from 2 mm to 4 mm) have a significant impact on the coupling between core A and core B. Table III summarizes the performance of rectangular-shape (this work) and circular-shape [19] dual-core fibers. In comparison with [19], the proposed dual-core fiber supports strict polarization-maintaining propagation so that the transmission would be robust. Moreover, we calculate the normalized core-spacing which is defined as $d/\lambda$ (where $d$ is the core-spacing and $\lambda$ is the central wavelength of the operation bandwidth). Thanks to the high birefringence, our work achieves narrower normalized core-spacing compared with the circular-shape fibers, implying that the rectangular-shape dual-core fibers can facilitate a denser integration of multiple cores. Furthermore, our work illustrates the high-efficient fiber excitation and connection with THz WR5.1-port devices using the 3D-printed flanges, yet additional adaptors and loads are needed for circular-shape fibers [19].

V. CONCLUSION

In summary, we have designed, 3D-printed, and measured densely integrated THz subwavelength rectangular dielectric dual-core fibers. Numerical simulations indicate that the fractional power in core of fiber mode has a dominant effect on inter-core coupling performance. For standardized WR 5.1 port, the excitation efficiency of *b-pol.* mode could be up to 62.8%. We employ a cost-efficient dielectric 3D printing technique to fabricate the dual-core fibers and polymer flange structures which offer solid integration between the fiber samples and the WR5.1 ports. Dual-core fibers with coupling region lengths of 3.5 cm, 5 cm, and 6.5 cm and core-spacings of 2 mm, 2.5 mm, and 4 mm are measured and analyzed. Experimental characterizations demonstrate that for our proposed subwavelength rectangular dielectric dual-core fibers with a core-spacing of 4 mm, the channel isolations between core A and core B are over 15 dB. In a nutshell, the proposed dual-core fiber holds enormous potentials for high-capacity, dense-integration THz space-division-multiplexing communication systems. Furthermore, we envision that our designs could be expanded to develop multi-core fibers, which would further enhance the transmission capacity.

REFERENCES

[1] W. Li *et al.*, "Photonic Terahertz Wireless Communication: Towards the Goal of High-Speed Kilometer-Level Transmission," *J. Lightw. Technol.*, vol. 42, no. 3, pp. 1159-1172, 2024.
[2] M. Zhu *et al.*, "Ultra-wideband fiber-THz-fiber seamless integration communication system toward 6G: architecture, key techniques, and testbed implementation," *Sci. China Inf. Sci.*, vol. 66, no. 1, p. 113301, 2023.
[3] G. Xu and M. Skorobogatiy, "Wired THz communications," *J. Infrared, Millimeter, Terahertz Waves,* vol. 43, no. 9, pp. 728-778, Sep. 2022.
[4] T. S. Rappaport *et al.*, "Wireless communications and applications above 100 GHz: Opportunities and challenges for 6G and beyond," *IEEE Access,* vol. 7, pp. 78729-78757, 2019.
[5] Z. Chen *et al.*, "A survey on terahertz communications," *China Commun.*, vol. 16, no. 2, pp. 1-35, 2019.
[6] S. Atakaramians, S. Afshar V, T. M. Monro, and D. Abbott, " Terahertz dielectric waveguides," *Adv. Opt. Photon.,* vol. 5, no. 2, pp. 169-215, 2013.
[7] X. Yu, M. Sugeta, Y. Yamagami, M. Fujita, and T. Nagatsuma, "Simultaneous low-loss and low-dispersion in a photonic-crystal waveguide for terahertz communications," *Appl. Phys. Exp,* vol. 12, no. 1, p. 012005, 2019.
[8] H. Li *et al.*, "Broadband Single-Mode Hybrid Photonic Crystal Waveguides for Terahertz Integration on a Chip," *Adv. Mater. Technol.,* vol. 5, no. 7, p. 2000117, 2020.
[9] H. Li, Y. Zhang, Y. Liu, and S. Atakaramians, "Terahertz Hybrid Topological Chip for 10-Gbps Full-Duplex Communications," *Electronics,* vol. 12, no. 1, p. 109, 2022.
[10] M. T. A. Khan, H. Li, N. N. M. Duong, A. Blanco-Redondo, and S. Atakaramians, "3D-Printed Terahertz Topological Waveguides," *Adv. Mater. Technol.,* vol. 6, no. 7, p. 2100252, 2021.
[11] W. Gao, X. Yu, M. Fujita, T. Nagatsuma, C. Fumeaux, and W. Withayachumnankul, "Effective-medium-cladded dielectric waveguides for terahertz waves," *Opt. Exp.,* vol. 27, no. 26, pp. 38721-38734, Dec. 2019.
[12] K. Nallappan, Y. Cao, G. Xu, H. Guerboukha, C. Nerguizian, and M. Skorobogatiy, "Dispersion-limited versus power-limited terahertz communication links using solid core subwavelength dielectric fibers," *Photon. Res.,* vol. 8, no. 11, pp. 1757-1775, 2020.
[13] Y. Wang, W. Gao, and C. Han, "End-to-end modeling and analysis for terahertz wireline transmission system with solid polymer fiber," in *ICC 2021-IEEE Int. Conf. on Commun.*, 2021, pp. 1-6.
[14] N. Van Thienen, Y. Zhang, M. De Wit, and P. Reynaert, "An 18Gbps polymer microwave fiber (PMF) communication link in 40nm CMOS," in *ESSCIRC Conf. 2016: 42nd Eur. Solid-State Circuits Conf.*, 2016, pp. 483-486.
[15] H. Li *et al.*, "Flexible single-mode hollow-core terahertz fiber with metamaterial cladding," *Optica,* vol. 3, no. 9, pp. 941-947, 2016.
[16] G. Xu and M. Skorobogatiy, "Continuous fabrication of polarization maintaining fibers via an annealing improved infinity additive manufacturing technique for THz communications," *Opt. Exp.,* vol. 31, no. 8, pp. 12894-12911, Apr. 2023.
[17] Y. Tan *et al.*, "Transmission of High-Frequency Terahertz Band Signal Beyond 300 GHz Over Metallic Hollow Core Fiber," *J. Lightw. Technol.,* vol. 40, no. 3, pp. 700-707, 2022.
[18] J. Ding *et al.*, "352-Gbit/s single line rate THz wired transmission based on PS-4096QAM employing hollow-core fiber," *Digit. Commun. Netw.,* vol. 9, no. 3, pp. 717-722, 2023.
[19] Y. Li, S. Liao, Q. Xue, and W. Che, "Coupling analysis of polytetrafluoroethylene dielectric waveguides in the sub-THz band," *Microw. Opt. Technol. Lett.,* vol. 65, no. 10, pp. 2691-2696, 2023.
[20] S. Li, H. Zhang, Y. Hou, J. Bai, W. Liu, and S. Chang, "Terahertz polarization splitter based on orthogonal microstructure dual-core photonic crystal fiber," *Appl. Opt.,* vol. 52, no. 14, pp. 3305-3310, May. 2013.
[21] H. Li *et al.*, "Terahertz polarization-maintaining subwavelength filters," *Opt. Exp.,* vol. 26, no. 20, pp. 25617-25629, Oct. 2018.
[22] H. Chen, G. Yan, E. Forsberg, and S. He, "Terahertz polarization splitter based on a dual-elliptical-core polymer fiber," *Appl. Opt.,* vol. 55, no. 23, pp. 6236-6242, Aug. 2016.
[23] V. Kumar, R. Varshney, and S. Kumar, "Design of a compact and broadband terahertz polarization splitter based on gradient dual-core photonic crystal fiber," *Appl. Opt.,* vol. 59, no. 7, pp. 1974-1979, Mar. 2020.
[24] J.-X. Zhang, "Dual-core PCF-based THz polarization beam splitter with broad bandwidth and ultra-high extinction ratio," *Optik,* vol. 251, p. 168425, 2022.
[25] Y. Zheng, J. Liu, Y. Wang, H. Liu, W. Wang, and X. Xie, "Design of compact dual-core terahertz polarization beam splitter with broad bandwidth," *Microw. Opt. Technol. Lett.,* vol. 65, no. 5, pp. 1277-1284, 2023.
[26] M. Weidenbach *et al.*, "3D printed dielectric rectangular waveguides, splitters and couplers for 120 GHz," *Opt. Exp.,* vol. 24, no. 25, pp. 28968-28976, Dec. 2016.
[27] J. Ma, M. Weidenbach, R. Guo, M. Koch, and D. M. Mittleman, "Communications with THz Waves: Switching Data Between Two Waveguides," *J. Infrared, Millimeter, Terahertz Waves,* vol. 38, no. 11, pp. 1316-1320, 2017.
[28] L. Jie, H. Li, Ya. Liu, J. Wang, G. Ren, and L. Pei, " Terahertz Multidimensional-Multiplexing and Refractive-Index-Sensing Integrated Device," *Acta Optica Sinica,* vol. 44, no. 8, p. 0823001, 2024.
[29] K. Nielsen, H. K. Rasmussen, P. U. Jepsen, and O. Bang, "Broadband terahertz fiber directional coupler," *Opt. Lett.,* vol. 35, no. 17, pp. 2879-2881, Sep. 2010.
[30] M.-Y. Chen, X.-X. Fu, and Y.-K. Zhang, "Design and analysis of a low-loss terahertz directional coupler based on three-core photonic crystal fibre configuration," *Journal of Physics D: Appl. Phys.,* vol. 44, no. 40, p. 405104, 2011.
[31] Y.-Y. Yu, X.-Y. Li, B. Sun, and K.-P. He, "Design and optimization of terahertz directional coupler based on hybrid-cladding hollow waveguide with low confinement loss," *Chinese Phys. B,* vol. 24, no. 6, p. 068702, 2015.
[32] Y.-F. Zhu *et al.*, "Low loss and polarization-insensitive coupling length for a terahertz fiber directional coupler with symmetric dual-suspended core structure," *Opt. Commun.,* vol. 480, p. 126497, 2021.
[33] S. Liu, J. Liu, Y. Li, and J. Zhang, "THz sensor based on dual-core PCF with defect core in detecting adulteration of olive oil," *Opt. Quantum Electro.,* vol. 53, no. 425, pp. 1-10, 2021.
[34] S. Li, H. Zhang, F. Fan, and S. Chang, "Terahertz ultrasensitive dual-core photonic crystal fiber microfluidic sensor for detecting high-absorption analytes," *Appl. Opt.,* vol. 60, no. 19, pp. 5716-5722, Jul. 2021.
[35] G. P. Mishra, D. Kumar, V. S. Chaudhary, and S. Sharma, "Terahertz refractive index sensor with high sensitivity based on two-core photonic crystal fiber," *Microw. Opt. Technol. Lett.,* vol. 63, no. 1, pp. 24-31, 2021.
[36] H. Li *et al.*, "Terahertz polarization-maintaining subwavelength dielectric waveguides," *J. Opt.,* vol. 20, no. 12, p. 125602, 2018.
[37] M. T. A. Khan, H. Li, Y. Liu, G.-D. Peng, and S. Atakaramians, "Compact terahertz birefringent gratings for dispersion compensation," *Opt. Exp.,* vol. 30, no. 6, pp. 8794-8803, Mar. 2022.
[38] COMSOL. (2023, Nov. 13). The COMSOL Product Suite [Online]. Available: https://cn.comsol.com/.
[39] K. Dorozhkin, D. Teterina, A. Badin, and V. Moskalenko, "ABS and PLA sub-terahertz absorbers for 3D-printing technology," in *J. Phys.: Conf. Ser.*, 2020, p. 012008.
[40] S. Zhang *et al.*, "Theoretical study of dual-core photonic crystal fibers with metal wire," *IEEE Photon. J.,* vol. 4, no. 4, pp. 1178-1187, 2012.
[41] Q. Wang, S. D. A. Shah, H. Li, B. Kuhlmey, and S. Atakaramians, "20 dB improvement utilizing custom-designed 3D-printed terahertz horn coupler," *Opt. Exp.,* vol. 31, no. 1, pp. 65-74, Jan. 2023.
[42] M. S. Islam, C. M. B. Cordeiro, M. A. R. Franco, J. Sultana, A. L. S. Cruz, and D. Abbott, "Terahertz optical fibers [Invited]," *Opt. Exp.,* vol. 28, no. 11, pp. 16089-16117, May. 2020.
[43] A. Araya *et al.*, "Absolute-length determination of a long-baseline Fabry–Perot cavity by means of resonating modulation sidebands," *Appl. Opt.,* vol. 38, no. 13, pp. 2848-2856, May. 1999.